\documentclass[aps,prl,twocolumn,superscriptaddress,reprint,showpacs]{revtex4-1}  
\usepackage{graphicx}
\usepackage{setspace}
\usepackage{amsmath}
\usepackage{amssymb}
\usepackage{color}
\usepackage{array}
\usepackage{subfigure}
\usepackage{hyperref}
\usepackage{float}
\usepackage{lipsum}

\usepackage[all]{xy}
\newcommand{\RN}[1]{%
  \textup{\uppercase\expandafter{\romannumeral#1}}%
}

\begin{document}

\title{Doppler-insensitive two-qubit controlled-PHASE gate protocol with dual-pulse off-resonant modulated driving for neutral atoms}

\author{Yuan Sun}
\email[email: ]{yuansun@live.com}
\affiliation{Key Laboratory of Quantum Optics and Center of Cold Atom Physics, Shanghai Institute of Optics and Fine Mechanics, Chinese Academy of Sciences, Shanghai 201800, China}
\author{Peng Xu}
\affiliation{State Key Laboratory of Magnetic Resonance and Atomic and Molecular Physics, Wuhan Institute of Physics and Mathematics, Chinese Academy of Sciences -- Wuhan National Laboratory for Optoelectronics, Wuhan 430071, P.R.China}
\affiliation{Center for Cold Atom Physics, Chinese Academy of Sciences, Wuhan 430071, P.R.China}
\author{Liang Liu}
\email[email: ]{liang.liu@siom.ac.cn}
\affiliation{Key Laboratory of Quantum Optics and Center of Cold Atom Physics, Shanghai Institute of Optics and Fine Mechanics, Chinese Academy of Sciences, Shanghai 201800, China}

\begin{abstract}
For neutral atom qubits, the residual thermal motion of the cold atoms constitutes a major challenge that limits the accessible two-qubit gate fidelity. Recently, an interesting type of two-qubit controlled-PHASE quantum gate protocol has been introduced for neutral atom qubit platform, which relies upon off-resonant modulated driving and Rydberg blockade effect. Building upon this progress, we have further developed an upgrade in the form of dual-pulse off-resonant modulated driving. Besides the inherent advantages of avoiding shelving population in Rydberg levels, not necessarily requiring individual site addressing, not sensitive to the exact value of blockade shift while suppressing population leakage error and rotation error, the major new feature of this protocol is Doppler-insensitive. In principle, the gate fidelity remains reasonably high over a relatively significant velocity range of the qubit atoms. Moreover, we anticipate that this protocol will inspire future improvements in quantum gate protocols for other types of qubit platforms, and its strategies may find applications in the area of quantum optimal control.
\end{abstract}
\maketitle



Cold neutral atoms in optical traps have long been deemed as an ideal choice of qubit platform, where Rydberg blockade \cite{PhysRevLett.85.2208, nphys1178, nphys1183, RevModPhys.82.2313, J.Phys.B.49.202001} serves as the backbone for two-qubit controlled-PHASE gate. Research effort in this area is not only important for quantum logic processing \cite{PhysRevLett.104.010503, PhysRevA.92.022336, PhysRevLett.119.160502}, but also crucial for quantum simulation \cite{nature24622} and quantum metrology \cite{nphoton.2011.35, RevModPhys.89.035002}. Recently, rapid progress has been made over a wide range of experimental topics in this area, including the increment of number of qubits in array format \cite{Saffman2019arXiv}, the enhancement of cat state size \cite{Omran570}, and the improvement of quantum gate performance \cite{Lukin2019arXiv}. The results so far clearly demonstrate the promising potential of neutral atom qubit and pave the way for its further development \cite{PhysRevA.66.065403, PhysRevLett.99.260501, PhysRevLett.107.093601, PhysRevLett.107.133602, PhysRevLett.109.233602, Dudin887Science, PhysRevLett.110.103001, PhysRevLett.113.053601, PhysRevA.92.022336, PhysRevLett.117.223001, PhysRevLett.119.160502, Hao2015srep, PhysRevA.93.040303, PhysRevA.93.041802, PhysRevA.94.053830, PhysRevA.95.041801, OPTICA.5.001492}. Amid many pressing tasks \cite{PhysRevLett.109.233602, PhysRevLett.112.040501, PhysRevLett.115.093601, PhysRevLett.119.113601, PhysRevA.91.030301, ISI:000457492900011}, an imminent next step for this platform is to enhance the two-qubit gate fidelity towards the goal of Noisy Intermediate-Scale Quantum technology (NISQ) under realistic experimental conditions.

Although intense attention and interest has been attracted \cite{PhysRevA.72.022347, PhysRevA.77.032723, RevModPhys.82.2313, PhysRevA.88.010303, PhysRevA.88.062337,  PhysRevA.91.012337, PhysRevA.92.042710, PhysRevA.94.032306, PhysRevA.96.042306}, currently the two-qubit controlled-PHASE gate fidelity is still below 99\% in experimental demonstrations, which poses a challenge for neutral atom qubit platform. Obstacles towards a higher fidelity can basically be divided into two categories: technical disturbances and intrinsic limitations. Technical disturbances include phase noise and amplitude fluctuations of laser pulses, errors in qubit state preparation and detection, atoms' relative positions within laser spatial profile and so on. Intrinsic limitations basically include spontaneous emissions from Rydberg levels, finite Rydberg blockade shift and residual thermal motion of the atoms. While technical disturbances call for further refinement in experimental and engineering techniques, experimental and theoretical progress over the last a few years has already paved the way for dealing with population leakage caused by Rydberg levels' decays and rotation error caused by finite Rydberg blockade shift. Therefore, the glaring issue of intrinsic limitations for now has become suppressing adverse effects caused by the residual thermal motion. In other words, a fast and robust two-qubit gate protocol with the purpose of remaining at reasonably high fidelity across a wide velocity range is in need.

Lately, a new category of efficient, robust and high-fidelity two-qubit controlled-PHASE gate protocols via the process of off-resonant modulated driving has been introduced \cite{Sun2018arXiv}. It resorts to specially tailored waveforms to gain appropriate phase accumulations under the presence of two-body dipole-dipole interaction. Besides the novelty of completing gate operation within a single sub-microsecond optical pulse, its features also include suppressing rotation error and avoiding shelved Rydberg population. Theoretical investigations show that it is more robust against Doppler induced dephasing compared with the commonly used multi-pulse protocol such as the $\pi$-gap-$\pi$ pulse sequence. However, for applications with high demand in fidelity, it sill puts a tight limit on the atom temperature such that sophisticated cooling mechanism towards motional ground state of optical trap will be required. Then, under the framework of off-resonant modulated driving, drastic changes are required in order to find a protocol with much enhanced tolerance for residual thermal motion of atomic qubits. This motivates us to investigate further along this direction.

In this letter, we report our recent progress in theoretically designing and analyzing a two-qubit controlled-PHASE gate protocol based upon Rydberg blockade for neutral atom platform, which aims at a generic reduction of susceptibility to atomic qubits' non-zero velocities. The enhancement in Doppler-insensitive properties comes mainly from a self-cancellation mechanism for adverse effects induced by finite velocities in our new design of dual-pulse technique. In the mean time, its generic characteristics also include avoiding shelving population in Rydberg levels, not necessarily requiring individual site addressing, not sensitive to the exact value of blockade shift and suppressing population leakage error and rotation error.

\begin{figure}[t]
\centering
\includegraphics[width=\linewidth]{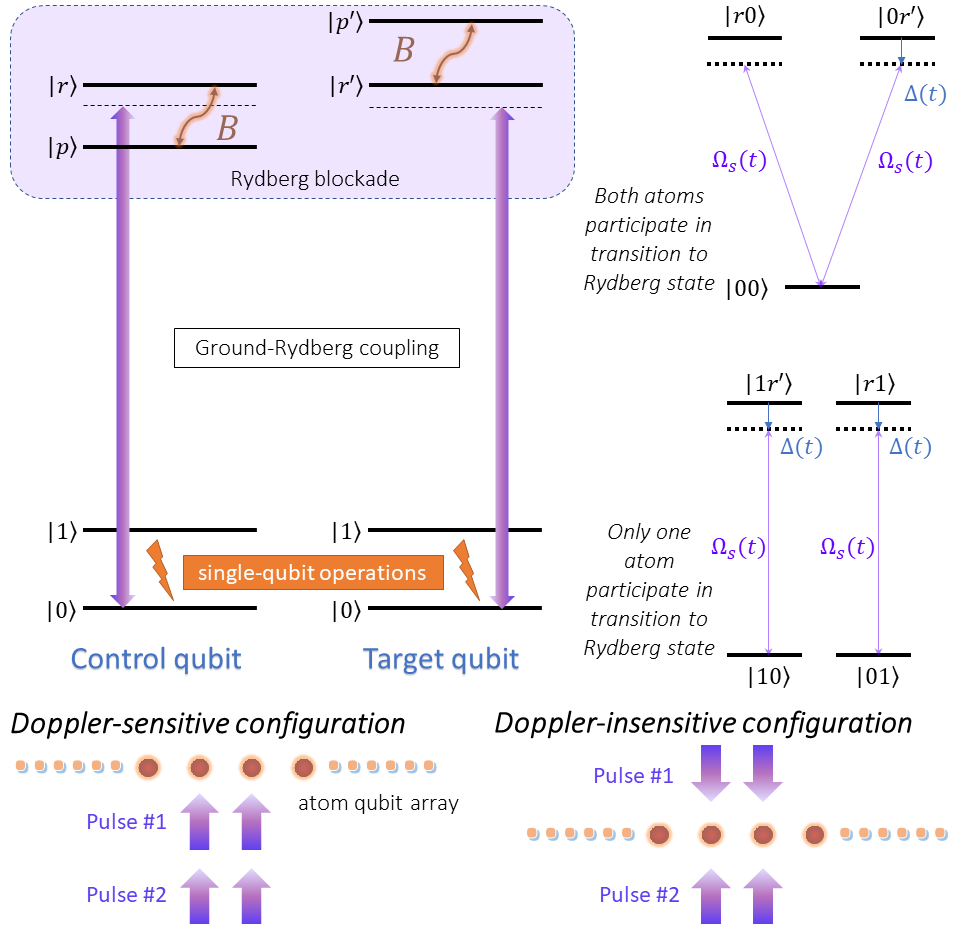}
\caption{Schematics of the Rydberg blockade phase gate under investigation. Top left: the relevant atomic states including the Rydberg blockade between $|r\rangle$ and $|r'\rangle$, where the lasers are driving $|0\rangle \leftrightarrow |r\rangle$ on control atom and $|0\rangle \leftrightarrow |r'\rangle$ on target atom. Top right: under ideal blockade situation, the linkage pattern for states participating the ground-Rydberg transitions $|01\rangle, |10\rangle$ and $|00\rangle$. Morris-Shore transform is also relevant in comprehending the linkage structures \cite{PhysRevA.27.906}. State $|11\rangle$ does not participate the prescribed interactions and stays unchanged through the process. Rydberg states $|r\rangle$ and $|r'\rangle$ may be the same or different, depending on the choice of F\"orster resonance structure. Bottom: two types of the dual-pulse driving configurations with atomic qubit array. Unlike previous proposals to appease Doppler-induced dephasing such as Ref. \cite{PhysRevA.91.012337}, it does not pose requirement on polarizations.}
\label{fig:basic_0}
\end{figure}

A sketch for relevant ingredients of the atom-light interaction is shown in Fig. \ref{fig:basic_0}. The qubit basis states of the atoms may be represented by a pair of long-lived hyperfine ground clock states for typical alkali atoms, which can be manipulated by external microwave field or optical stimulated Raman transition \cite{PhysRevLett.114.100503, PhysRevA.92.022336, J.Phys.B.49.202001}. Modulated laser pulses will be applied to drive the ground-Rydberg transitions of the control and target atoms. In particular, this protocol employs two consecutive pulses to realize the performance of controlled-PHASE gate. When combined with a local Hadamard gate on the target qubit atom ($\pi/2$ rotation for transition $|0\rangle \leftrightarrow |1\rangle$), this leads to the universal controlled-NOT gate \cite{PhysRevLett.85.2208, nphys1178, RevModPhys.82.2313, J.Phys.B.49.202001}. If $|r\rangle, |r'\rangle$ are the same, then individual site addressing is not mandatory and the experiment can be operated through one global laser. For simplicity, throughout this article the condition of symmetric driving will be presumed, namely both the qubit atoms will receive the same Rabi frequency and detuning in their effective ground-Rydberg transition couplings. 

Under these prescribed conditions, for $|01\rangle$ and $|10\rangle$, the situation reduces to a two-level system of ground-Rydberg transition with time-dependent Rabi frequency $\Omega_s(t)$ and a fixed detuning $\Delta$. Meanwhile, the dynamics of $|00\rangle$ involves the Rydberg dipole-dipole interaction with linkage structure as $|00\rangle \leftrightarrow |R\rangle \leftrightarrow |rr'\rangle \leftrightarrow |pp'\rangle$ where we define $|R\rangle = (|r0\rangle+|0r'\rangle)/\sqrt{2}$. The interaction Hamiltonian is then:
\begin{eqnarray}
\label{eq:Rydberg_Hamiltonian}
H_I/\hbar = &\frac{\sqrt{2}}{2}\Omega_s|R \rangle \langle 00| + \frac{\sqrt{2}}{2}\Omega_s|rr'\rangle \langle R| + \text{H.c.}
\nonumber\\
&+ \Delta |R\rangle\langle R| + 2\Delta |rr'\rangle\langle rr'|, 
\end{eqnarray}
where rotating wave approximation is already included. For the F\"oster resonance structure of $|rr'\rangle \leftrightarrow |pp'\rangle$, we set the coupling strength as $B$ and the small F\"{o}rster energy penalty term as $\delta_p$ for $|pp'\rangle$:
\begin{equation}
\label{eq:blockade_Hamiltonian}
H_F/\hbar = B|pp'\rangle \langle rr'| + \text{H.c.} 
+ \delta_p |pp'\rangle\langle pp'|,
\end{equation}
such that the overall Hamiltonian is $H=H_I + H_F$.

Under the assumption of ideal Rydberg blockade where double Rydberg excitation into $|rr'\rangle$ is forbidden, there exist two categories of couplings: $|10\rangle\leftrightarrow|1r'\rangle$, $|01\rangle\leftrightarrow|r1\rangle$ with Rabi frequency $\Omega_s$ and $|00\rangle \leftrightarrow |R\rangle$ with Rabi frequency $\sqrt{2}\Omega_s$, as can be seen in the linkage structure of Fig. \ref{fig:basic_0}. For our case of dual-pulse gate operation, we are interested in that the two pulses are identical and the waveform receives only amplitude modulation, to avoid the experimental complexities of hybrid modulation of both amplitude and frequency. Constraints for an appropriate waveform include: (1) the population returns to ground state with unity probability after each pulse for both categories of $\Omega_s, \sqrt{2}\Omega_s$; (2) the accumulated phases satisfy the general controlled-PHASE gate condition after the entire interaction. More specifically, the phase constraint after interacting with one modulated pulse is:
\begin{equation}
\label{phase_constraint}
2\phi_{11} = \pm \pi - 2\phi_{00} + 2\phi_{01} + 2\phi_{10},
\end{equation}
which reduces to $2(\phi_{01} + \phi_{10} - \phi_{00}) = \pm\pi$ if $\phi_{11} = 0$.

\begin{figure}[t]
\centering
\fbox{\includegraphics[width=\linewidth]{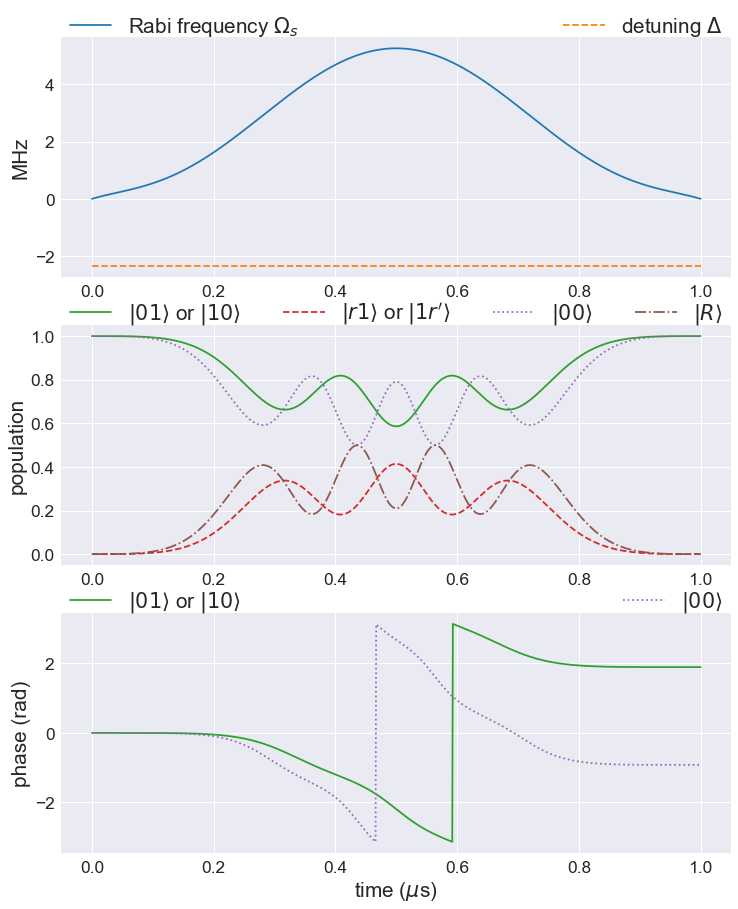}}
\caption{Numerical simulation of the time evolution. The waveform is set according to Eq. \eqref{eq:waveform_v22g}, while $B=2\pi \times 500$ MHz, $\delta_p = 2\pi \times -3$ MHz. The first graph shows the waveform, the second graph shows the population on different atomic states, while the last graph shows the phase accumulation of the atomic wave function during the process. This calculation serves for 'pinned-down' atoms with zero velocity.}
\label{fig:single_pulse}
\end{figure}

The essential point is to represent the continuous waveform via a set of discrete parameters. For this purpose, Fourier series and Bernstein polynomials float as attractive candidates, and from mathematical point of view, either choice form a complete basis for non-pathological functions defined on a finite time interval. Nevertheless, a limited amount of terms in the expansion will be enough for the practical computing demands of designing pulse waveform. With the purpose of amplitude modulation, we choose to proceed with Bernstein polynomials defined on time interval $[0, T_g]$. For ease of experimental implementations, preferably the amplitude-modulated pulse shall start and end at zero intensity. Let $b_{\nu, n}$ denote the $\nu$th Bernstein basis polynomial of degree $n$, the abstract form of the pursued pulse is described below:
\begin{equation}
\label{eq:waveform_v22g}
\Omega_s (t) = \sum_{\nu=1}^{n-1} \beta_\nu b_{\nu, n}(t/T_g),\, 
\Delta (t) = \Delta_0 \equiv \textit{ constant}.
\end{equation}

Indeed, appropriate solutions can be obtained. Basically, first we sketch waveforms under the assumption of perfect adiabatic evolution, and then carry out numerical optimization procedures for refinement towards desired high-fidelity \cite{Sun2018arXiv, SuppInfo}. As a sample calculation, we set $n=8$ in Eq. \eqref{eq:waveform_v22g} and intentionally seek a symmetric waveform. For pulse time $T_g$ set as 1 $\mu$s, we have reached a set of satisfying parameters, $\beta_1=\beta_7=0.794, \beta_2=\beta_6=0, \beta_3=\beta_5=5.841, \beta_4=9.725, \Delta_0=-2.360$; all coefficient units are MHz. The result for interaction process with one pulse is shown in Fig. \ref{fig:single_pulse}. This corresponds to single-photon uv transition for typical alkali atoms.

\begin{figure}[b]
\centering
\fbox{\includegraphics[width=\linewidth]{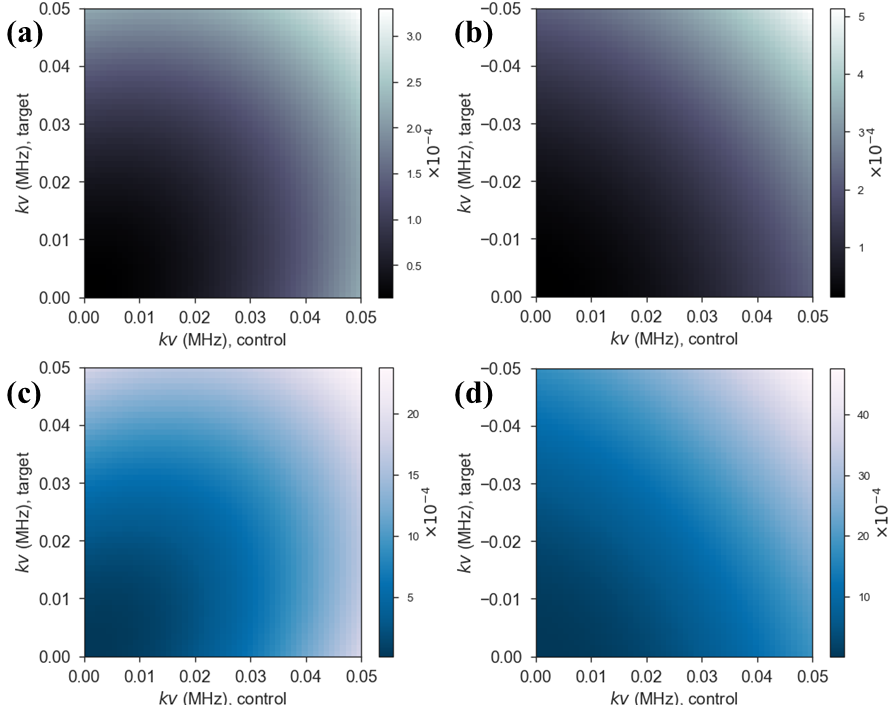}}
\caption{Numerical simulation for gate error $\mathcal{E}$ with atomic qubits of non-zero velocities, with dual-pulse optical driving according to the waveform of Fig. \ref{fig:single_pulse}. Spontaneous emission of Rydberg states is neglected. (a) (b): counter-propagating configuration, Doppler-insensitive; (c) (d): co-propagating configuration, Doppler-sensitive.}
\label{fig:velocity_scan}
\end{figure}

Interaction with two consecutive pulses as Fig. \ref{fig:single_pulse} will make a controlled-PHASE gate. In principle, since population returns to ground state, the gap time between two pulses does not matter and no extra phase accumulation will be induced. Although overall it requires two pulses, there exists no demand on the directions where the pulse come from. For example, two opposite choices may be employed: co-propagating and counter-propagating as drawn in Fig. \ref{fig:basic_0}. Effectively, atoms with different velocities see different detunings of the optical driving pulse as a consequence of Doppler shift, and this is exactly the cause of Doppler-dephasing problem in previously known multi-pulse Rydberg blockade controlled-PHASE gate protocols. In order to investigate such effects caused by cold qubit atoms' residual thermal motions, we compute the gate performance under different atom velocities and the result is shown in Fig. \ref{fig:velocity_scan}. The fidelity is calculated as $F = ( \text{Tr}(MM^\dagger) + |\text{Tr}(M)|^2 )/20$ \cite{PhysRevA.94.032306, PhysRevA.96.042306}, where $M = U_\text{C-Z}^\dagger U$ with $U_\text{C-Z}, U$ being the transform matrices of an ideal C-Z gate and our gate protocol after local phase rotation respectively. Then the gate error is $\mathcal{E} = 1- F$. 

From Fig. \ref{fig:velocity_scan}, we observe that in the counter-propagating configuration, the adverse influence caused by residual thermal motion is significantly suppressed. In the evaluation process, the atomic motion is treated classically with finite prescribed velocities, which do not change after interacting with the driving pulse since the atom returns to ground state. For the ground-Rydberg transition with $|01\rangle$ or $|10\rangle$, the situation is similar to the case of two level atom driven by constant Rabi frequency and detuning with comparable magnitudes. Namely, the first order effect is a phase shift $\sim kvT_g$ on the ground state, with $k$ being the wave vector of the driving laser. Other adverse effects belong to the category of second order corrections, such as residual population on the Rydberg states. Therefore, after interacting with the second pulse of an opposite wave vector $-k$, the first order phase shift cancels out automatically. The case of $|00\rangle$ is slightly more complicated since the atomic motions break the energy degeneracy. The first order effect of phase shift accumulated on the ground level contains contributions from motions of both the control and target qubit atoms. Nevertheless, the second counter-propagating pulse is still effective in reducing the overall error. Roughly speaking, the counter-propagating dual-pulse driving technique invokes a self-cancellation mechanism and suppresses the Doppler-induced adverse effects to second order of $\sim kvT_g$ and therefore we regard it as Doppler-insensitive. 

\begin{figure}[th]
\centering
\includegraphics[width=0.95\linewidth]{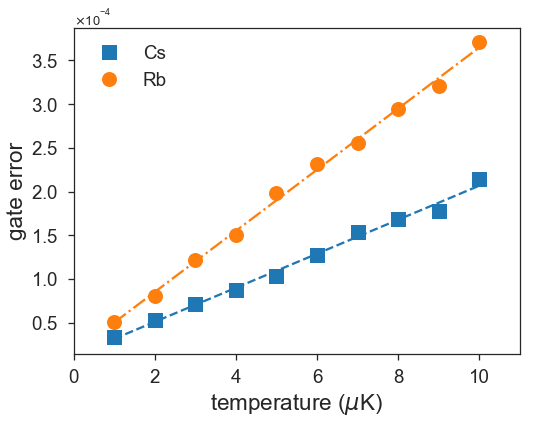}
\caption{Numerical simulation for gate error $\mathcal{E}$ with respect to different temperature settings, via Monte-Carlo approaches. Spontaneous emission of Rydberg states is neglected. Fittings to straight lines are included.}
\label{fig:fidelity_temperature_scan}
\end{figure}

Next, we proceed to evaluate the performance of this dual-pulse technique for typical alkali atoms such as Rb and Cs. Without loss of generality, we assume that the atoms obey 3D Maxwell-Boltzmann distribution and apply a Monte-Carlo approach to sample the velocity component along the optical axis of driving lasers. The numerical result is shown in Fig. \ref{fig:fidelity_temperature_scan} with respect to the waveform of Fig. \ref{fig:single_pulse}. We observe that the reduction of fidelity is well controlled on the order of $10^{-4}$ with 1 $\mu$s duration pulse for typical cold atom temperatures, significantly higher above the recoil limit temperature.

\begin{figure}[th]
\centering
\includegraphics[width=0.95\linewidth]{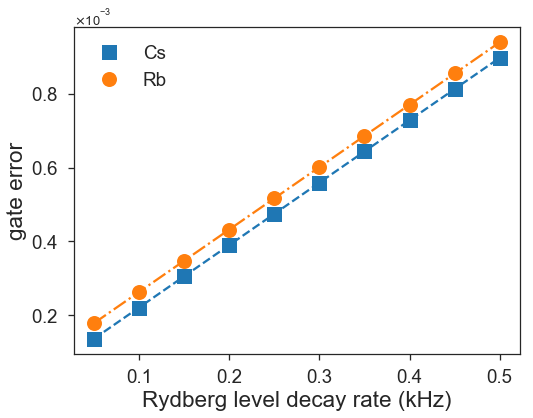}
\caption{Numerical simulation for gate error $\mathcal{E}$ with respect to different Rydberg decay rates via Monte-Carlo approaches, with temperature set as 2 $\mu$K. For simplicity, the spontaneous decay rates of all Rydberg states are taken as the same. Fittings to straight lines are included.}
\label{fig:fidelity_decay_scan}
\end{figure}

For dynamics associated with $|00\rangle$, the singly-excited Rydberg state $|R\rangle$ does not receive heavy population throughout the interaction process. It draws a clear difference compared with on-resonance quantum Rabi oscillation, while sharing similarities with adiabatic rapid passage. Moreover, the mechanism of adiabatically tracking the two-atom dark state mostly composing of $|R\rangle$ and $|pp'\rangle$ plays an important role here, where the smooth waveform of amplitude modulation helps to suppress the rotation error and keep a minimal population on $|rr'\rangle$. Apparently, it does not require the exact knowledge of Rydberg blockade shift strength. In other words, we anticipate that major limitations on the attainable fidelity mostly comes from spontaneous emissions, modulation imperfections and technical noises. Therefore, provided the Rydberg blockade shift is strong enough, we deduce that in principle the Rydberg levels' spontaneous emission is the dominating theoretical limiting factor towards achieving high-fidelity. To estimate its influences, we compute gate error as a function of the Rydberg decay rates and the result is shown in Fig. \ref{fig:fidelity_decay_scan}.

So far, we've been focusing on one-photon transition for the ground-Rydberg coupling, typically driven by uv laser. On the other hand, two-photon transition with large one-photon detuning by two driving lasers of longer wavelength has been a well established subject, and our protocol is also compatible with such arrangements. However, the situation becomes more complicated due to various sources of extra ac Stark shifts and decoherences \cite{PhysRevA.72.022347, PhysRevA.94.032306}. 

We think that this protocol is well compatible with the currently available mainstream hardware of neutral atom qubit platforms \cite{PhysRevA.92.022336, PhysRevLett.119.160502}, and we are looking forward to experimental instantiation in near future. Under parameters from realistic experimental apparatus, this protocol may be designed for fast gate operation below 1 $\mu$s \cite{SuppInfo}. Carrying out entangling gate within smooth and continuous shaped driving pulse has already become a common practice in various qubit platforms \cite{PhysRevLett.107.080502, PhysRevA.94.032306}, and we think the introduction of the concept of non-trivial Doppler-insensitive mechanism into this area will inspire further updates and improvements. We also anticipate that our work will help the efforts in the ensemble qubit approach \cite{PhysRevLett.115.093601, PhysRevLett.119.180504} and the Rydberg-mediated atom-photon controlled-PHASE gate \cite{Hao2015srep, PhysRevA.93.040303, PhysRevA.94.053830, OPTICA.5.001492}. 

In conclusion, we have designed and analyzed a new category of two-qubit controlled-PHASE gate based upon Rydberg blockade effects, with the underlying technique of dual-pulse off-resonant modulated optical driving. While avoiding shelving population on Rydberg states in gap time, the essential feature is that the gate fidelity becomes much more robust against the qubit atoms' thermal motion, as a result of the self-cancellation mechanism for the Doppler-induced phase disturbance. The gate fidelity is not sensitive to the exact value of blockade shift and rotation error is suppressed, thanks to the dark state driving mechanism. In theory, the major intrinsic error source comes from the spontaneous emission from Rydberg states. We not only demonstrate the mechanisms of the protocol, but also discuss the principles behind the process of constructing appropriate ingredients for the protocol. Our aim with the Rydberg blockade gate is that high-quality ground-Rydberg Rabi coherence shall be directly translated into high-fidelity controlled-PHASE gate, and future improvements will include efforts towards a faster gate operation, reduction of the population leakage error, and a more user-friendly parameter setting of the waveform.

\begin{acknowledgements}
The authors gratefully acknowledge the funding support from the National Key R\&D Program of China (under contract Grant No. 2016YFA0301504 and No. 2016YFA0302800). The authors gratefully thank the help from Professor Mark Saffman and Professor Mingsheng Zhan who essentially make this work possible. The authors also thank Professor Xiaodong He, Professor Dongsheng Ding and Professor Tian Xia for enlightening discussions.
\end{acknowledgements}

\bibliographystyle{apsrev4-1}

\renewcommand{\baselinestretch}{1}
\normalsize

\bibliography{quartette_ref}

\begin{thebibliography}{48}%
\makeatletter
\providecommand \@ifxundefined [1]{%
 \@ifx{#1\undefined}
}%
\providecommand \@ifnum [1]{%
 \ifnum #1\expandafter \@firstoftwo
 \else \expandafter \@secondoftwo
 \fi
}%
\providecommand \@ifx [1]{%
 \ifx #1\expandafter \@firstoftwo
 \else \expandafter \@secondoftwo
 \fi
}%
\providecommand \natexlab [1]{#1}%
\providecommand \enquote  [1]{``#1''}%
\providecommand \bibnamefont  [1]{#1}%
\providecommand \bibfnamefont [1]{#1}%
\providecommand \citenamefont [1]{#1}%
\providecommand \href@noop [0]{\@secondoftwo}%
\providecommand \href [0]{\begingroup \@sanitize@url \@href}%
\providecommand \@href[1]{\@@startlink{#1}\@@href}%
\providecommand \@@href[1]{\endgroup#1\@@endlink}%
\providecommand \@sanitize@url [0]{\catcode `\\12\catcode `\$12\catcode
  `\&12\catcode `\#12\catcode `\^12\catcode `\_12\catcode `\%12\relax}%
\providecommand \@@startlink[1]{}%
\providecommand \@@endlink[0]{}%
\providecommand \url  [0]{\begingroup\@sanitize@url \@url }%
\providecommand \@url [1]{\endgroup\@href {#1}{\urlprefix }}%
\providecommand \urlprefix  [0]{URL }%
\providecommand \Eprint [0]{\href }%
\providecommand \doibase [0]{https://doi.org/}%
\providecommand \selectlanguage [0]{\@gobble}%
\providecommand \bibinfo  [0]{\@secondoftwo}%
\providecommand \bibfield  [0]{\@secondoftwo}%
\providecommand \translation [1]{[#1]}%
\providecommand \BibitemOpen [0]{}%
\providecommand \bibitemStop [0]{}%
\providecommand \bibitemNoStop [0]{.\EOS\space}%
\providecommand \EOS [0]{\spacefactor3000\relax}%
\providecommand \BibitemShut  [1]{\csname bibitem#1\endcsname}%
\let\auto@bib@innerbib\@empty
\bibitem [{\citenamefont {Jaksch}\ \emph {et~al.}(2000)\citenamefont {Jaksch},
  \citenamefont {Cirac}, \citenamefont {Zoller}, \citenamefont {Rolston},
  \citenamefont {C\^ot\'e},\ and\ \citenamefont {Lukin}}]{PhysRevLett.85.2208}%
  \BibitemOpen
  \bibfield  {author} {\bibinfo {author} {\bibfnamefont {D.}~\bibnamefont
  {Jaksch}}, \bibinfo {author} {\bibfnamefont {J.~I.}\ \bibnamefont {Cirac}},
  \bibinfo {author} {\bibfnamefont {P.}~\bibnamefont {Zoller}}, \bibinfo
  {author} {\bibfnamefont {S.~L.}\ \bibnamefont {Rolston}}, \bibinfo {author}
  {\bibfnamefont {R.}~\bibnamefont {C\^ot\'e}},\ and\ \bibinfo {author}
  {\bibfnamefont {M.~D.}\ \bibnamefont {Lukin}},\ }\href
  {https://doi.org/10.1103/PhysRevLett.85.2208} {\bibfield  {journal} {\bibinfo
   {journal} {Phys. Rev. Lett.}\ }\textbf {\bibinfo {volume} {85}},\ \bibinfo
  {pages} {2208} (\bibinfo {year} {2000})}\BibitemShut {NoStop}%
\bibitem [{\citenamefont {Urban}\ \emph {et~al.}(2009)\citenamefont {Urban},
  \citenamefont {Johnson}, \citenamefont {Henage}, \citenamefont {Isenhower},
  \citenamefont {Yavuz}, \citenamefont {Walker},\ and\ \citenamefont
  {Saffman}}]{nphys1178}%
  \BibitemOpen
  \bibfield  {author} {\bibinfo {author} {\bibfnamefont {E.}~\bibnamefont
  {Urban}}, \bibinfo {author} {\bibfnamefont {T.~A.}\ \bibnamefont {Johnson}},
  \bibinfo {author} {\bibfnamefont {T.}~\bibnamefont {Henage}}, \bibinfo
  {author} {\bibfnamefont {L.}~\bibnamefont {Isenhower}}, \bibinfo {author}
  {\bibfnamefont {D.~D.}\ \bibnamefont {Yavuz}}, \bibinfo {author}
  {\bibfnamefont {T.~G.}\ \bibnamefont {Walker}},\ and\ \bibinfo {author}
  {\bibfnamefont {M.}~\bibnamefont {Saffman}},\ }\href
  {https://doi.org/10.1038/nphys1178} {\bibfield  {journal} {\bibinfo
  {journal} {Nature Physics}\ }\textbf {\bibinfo {volume} {5}},\ \bibinfo
  {pages} {110} (\bibinfo {year} {2009})}\BibitemShut {NoStop}%
\bibitem [{\citenamefont {Gaëtan}\ \emph {et~al.}(2009)\citenamefont
  {Gaëtan}, \citenamefont {Miroshnychenko}, \citenamefont {Wilk},
  \citenamefont {Chotia}, \citenamefont {Viteau}, \citenamefont {Comparat},
  \citenamefont {Pillet}, \citenamefont {Browaeys},\ and\ \citenamefont
  {Grangier}}]{nphys1183}%
  \BibitemOpen
  \bibfield  {author} {\bibinfo {author} {\bibfnamefont {A.}~\bibnamefont
  {Gaëtan}}, \bibinfo {author} {\bibfnamefont {Y.}~\bibnamefont
  {Miroshnychenko}}, \bibinfo {author} {\bibfnamefont {T.}~\bibnamefont
  {Wilk}}, \bibinfo {author} {\bibfnamefont {A.}~\bibnamefont {Chotia}},
  \bibinfo {author} {\bibfnamefont {M.}~\bibnamefont {Viteau}}, \bibinfo
  {author} {\bibfnamefont {D.}~\bibnamefont {Comparat}}, \bibinfo {author}
  {\bibfnamefont {P.}~\bibnamefont {Pillet}}, \bibinfo {author} {\bibfnamefont
  {A.}~\bibnamefont {Browaeys}},\ and\ \bibinfo {author} {\bibfnamefont
  {P.}~\bibnamefont {Grangier}},\ }\href {https://doi.org/10.1038/nphys1183}
  {\bibfield  {journal} {\bibinfo  {journal} {Nature Physics}\ }\textbf
  {\bibinfo {volume} {5}},\ \bibinfo {pages} {115} (\bibinfo {year}
  {2009})}\BibitemShut {NoStop}%
\bibitem [{\citenamefont {Saffman}\ \emph {et~al.}(2010)\citenamefont
  {Saffman}, \citenamefont {Walker},\ and\ \citenamefont
  {M\o{}lmer}}]{RevModPhys.82.2313}%
  \BibitemOpen
  \bibfield  {author} {\bibinfo {author} {\bibfnamefont {M.}~\bibnamefont
  {Saffman}}, \bibinfo {author} {\bibfnamefont {T.~G.}\ \bibnamefont
  {Walker}},\ and\ \bibinfo {author} {\bibfnamefont {K.}~\bibnamefont
  {M\o{}lmer}},\ }\href {https://doi.org/10.1103/RevModPhys.82.2313} {\bibfield
   {journal} {\bibinfo  {journal} {Rev. Mod. Phys.}\ }\textbf {\bibinfo
  {volume} {82}},\ \bibinfo {pages} {2313} (\bibinfo {year}
  {2010})}\BibitemShut {NoStop}%
\bibitem [{\citenamefont {Saffman}(2016)}]{J.Phys.B.49.202001}%
  \BibitemOpen
  \bibfield  {author} {\bibinfo {author} {\bibfnamefont {M.}~\bibnamefont
  {Saffman}},\ }\href {http://stacks.iop.org/0953-4075/49/i=20/a=202001}
  {\bibfield  {journal} {\bibinfo  {journal} {Journal of Physics B: Atomic,
  Molecular and Optical Physics}\ }\textbf {\bibinfo {volume} {49}},\ \bibinfo
  {pages} {202001} (\bibinfo {year} {2016})}\BibitemShut {NoStop}%
\bibitem [{\citenamefont {Isenhower}\ \emph {et~al.}(2010)\citenamefont
  {Isenhower}, \citenamefont {Urban}, \citenamefont {Zhang}, \citenamefont
  {Gill}, \citenamefont {Henage}, \citenamefont {Johnson}, \citenamefont
  {Walker},\ and\ \citenamefont {Saffman}}]{PhysRevLett.104.010503}%
  \BibitemOpen
  \bibfield  {author} {\bibinfo {author} {\bibfnamefont {L.}~\bibnamefont
  {Isenhower}}, \bibinfo {author} {\bibfnamefont {E.}~\bibnamefont {Urban}},
  \bibinfo {author} {\bibfnamefont {X.~L.}\ \bibnamefont {Zhang}}, \bibinfo
  {author} {\bibfnamefont {A.~T.}\ \bibnamefont {Gill}}, \bibinfo {author}
  {\bibfnamefont {T.}~\bibnamefont {Henage}}, \bibinfo {author} {\bibfnamefont
  {T.~A.}\ \bibnamefont {Johnson}}, \bibinfo {author} {\bibfnamefont {T.~G.}\
  \bibnamefont {Walker}},\ and\ \bibinfo {author} {\bibfnamefont
  {M.}~\bibnamefont {Saffman}},\ }\href
  {https://doi.org/10.1103/PhysRevLett.104.010503} {\bibfield  {journal}
  {\bibinfo  {journal} {Phys. Rev. Lett.}\ }\textbf {\bibinfo {volume} {104}},\
  \bibinfo {pages} {010503} (\bibinfo {year} {2010})}\BibitemShut {NoStop}%
\bibitem [{\citenamefont {Maller}\ \emph {et~al.}(2015)\citenamefont {Maller},
  \citenamefont {Lichtman}, \citenamefont {Xia}, \citenamefont {Sun},
  \citenamefont {Piotrowicz}, \citenamefont {Carr}, \citenamefont {Isenhower},\
  and\ \citenamefont {Saffman}}]{PhysRevA.92.022336}%
  \BibitemOpen
  \bibfield  {author} {\bibinfo {author} {\bibfnamefont {K.~M.}\ \bibnamefont
  {Maller}}, \bibinfo {author} {\bibfnamefont {M.~T.}\ \bibnamefont
  {Lichtman}}, \bibinfo {author} {\bibfnamefont {T.}~\bibnamefont {Xia}},
  \bibinfo {author} {\bibfnamefont {Y.}~\bibnamefont {Sun}}, \bibinfo {author}
  {\bibfnamefont {M.~J.}\ \bibnamefont {Piotrowicz}}, \bibinfo {author}
  {\bibfnamefont {A.~W.}\ \bibnamefont {Carr}}, \bibinfo {author}
  {\bibfnamefont {L.}~\bibnamefont {Isenhower}},\ and\ \bibinfo {author}
  {\bibfnamefont {M.}~\bibnamefont {Saffman}},\ }\href
  {https://doi.org/10.1103/PhysRevA.92.022336} {\bibfield  {journal} {\bibinfo
  {journal} {Phys. Rev. A}\ }\textbf {\bibinfo {volume} {92}},\ \bibinfo
  {pages} {022336} (\bibinfo {year} {2015})}\BibitemShut {NoStop}%
\bibitem [{\citenamefont {Zeng}\ \emph {et~al.}(2017)\citenamefont {Zeng},
  \citenamefont {Xu}, \citenamefont {He}, \citenamefont {Liu}, \citenamefont
  {Liu}, \citenamefont {Wang}, \citenamefont {Papoular}, \citenamefont
  {Shlyapnikov},\ and\ \citenamefont {Zhan}}]{PhysRevLett.119.160502}%
  \BibitemOpen
  \bibfield  {author} {\bibinfo {author} {\bibfnamefont {Y.}~\bibnamefont
  {Zeng}}, \bibinfo {author} {\bibfnamefont {P.}~\bibnamefont {Xu}}, \bibinfo
  {author} {\bibfnamefont {X.}~\bibnamefont {He}}, \bibinfo {author}
  {\bibfnamefont {Y.}~\bibnamefont {Liu}}, \bibinfo {author} {\bibfnamefont
  {M.}~\bibnamefont {Liu}}, \bibinfo {author} {\bibfnamefont {J.}~\bibnamefont
  {Wang}}, \bibinfo {author} {\bibfnamefont {D.~J.}\ \bibnamefont {Papoular}},
  \bibinfo {author} {\bibfnamefont {G.~V.}\ \bibnamefont {Shlyapnikov}},\ and\
  \bibinfo {author} {\bibfnamefont {M.}~\bibnamefont {Zhan}},\ }\href
  {https://doi.org/10.1103/PhysRevLett.119.160502} {\bibfield  {journal}
  {\bibinfo  {journal} {Phys. Rev. Lett.}\ }\textbf {\bibinfo {volume} {119}},\
  \bibinfo {pages} {160502} (\bibinfo {year} {2017})}\BibitemShut {NoStop}%
\bibitem [{\citenamefont {Bernien}\ \emph {et~al.}(2017)\citenamefont
  {Bernien}, \citenamefont {Schwartz}, \citenamefont {Keesling}, \citenamefont
  {Levine}, \citenamefont {Omran}, \citenamefont {Pichler}, \citenamefont
  {Choi}, \citenamefont {Zibrov}, \citenamefont {Endres}, \citenamefont
  {Greiner}, \citenamefont {Vuleti\'c},\ and\ \citenamefont
  {Lukin}}]{nature24622}%
  \BibitemOpen
  \bibfield  {author} {\bibinfo {author} {\bibfnamefont {H.}~\bibnamefont
  {Bernien}}, \bibinfo {author} {\bibfnamefont {S.}~\bibnamefont {Schwartz}},
  \bibinfo {author} {\bibfnamefont {A.}~\bibnamefont {Keesling}}, \bibinfo
  {author} {\bibfnamefont {H.}~\bibnamefont {Levine}}, \bibinfo {author}
  {\bibfnamefont {A.}~\bibnamefont {Omran}}, \bibinfo {author} {\bibfnamefont
  {H.}~\bibnamefont {Pichler}}, \bibinfo {author} {\bibfnamefont
  {S.}~\bibnamefont {Choi}}, \bibinfo {author} {\bibfnamefont {A.~S.}\
  \bibnamefont {Zibrov}}, \bibinfo {author} {\bibfnamefont {M.}~\bibnamefont
  {Endres}}, \bibinfo {author} {\bibfnamefont {M.}~\bibnamefont {Greiner}},
  \bibinfo {author} {\bibfnamefont {V.}~\bibnamefont {Vuleti\'c}},\ and\
  \bibinfo {author} {\bibfnamefont {M.~D.}\ \bibnamefont {Lukin}},\ }\href
  {https://doi.org/10.1038/nature24622} {\bibfield  {journal} {\bibinfo
  {journal} {Nature}\ }\textbf {\bibinfo {volume} {551}},\ \bibinfo {pages}
  {579} (\bibinfo {year} {2017})}\BibitemShut {NoStop}%
\bibitem [{\citenamefont {Giovannetti}\ \emph {et~al.}(2011)\citenamefont
  {Giovannetti}, \citenamefont {Lloyd},\ and\ \citenamefont
  {Maccone}}]{nphoton.2011.35}%
  \BibitemOpen
  \bibfield  {author} {\bibinfo {author} {\bibfnamefont {V.}~\bibnamefont
  {Giovannetti}}, \bibinfo {author} {\bibfnamefont {S.}~\bibnamefont {Lloyd}},\
  and\ \bibinfo {author} {\bibfnamefont {L.}~\bibnamefont {Maccone}},\ }\href
  {https://doi.org/10.1038/nphoton.2011.35} {\bibfield  {journal} {\bibinfo
  {journal} {Nature Photonics}\ }\textbf {\bibinfo {volume} {5}},\ \bibinfo
  {pages} {222} (\bibinfo {year} {2011})}\BibitemShut {NoStop}%
\bibitem [{\citenamefont {Degen}\ \emph {et~al.}(2017)\citenamefont {Degen},
  \citenamefont {Reinhard},\ and\ \citenamefont
  {Cappellaro}}]{RevModPhys.89.035002}%
  \BibitemOpen
  \bibfield  {author} {\bibinfo {author} {\bibfnamefont {C.~L.}\ \bibnamefont
  {Degen}}, \bibinfo {author} {\bibfnamefont {F.}~\bibnamefont {Reinhard}},\
  and\ \bibinfo {author} {\bibfnamefont {P.}~\bibnamefont {Cappellaro}},\
  }\href {https://doi.org/10.1103/RevModPhys.89.035002} {\bibfield  {journal}
  {\bibinfo  {journal} {Rev. Mod. Phys.}\ }\textbf {\bibinfo {volume} {89}},\
  \bibinfo {pages} {035002} (\bibinfo {year} {2017})}\BibitemShut {NoStop}%
\bibitem [{\citenamefont {Graham}\ \emph {et~al.}()\citenamefont {Graham},
  \citenamefont {Kwon}, \citenamefont {Grinkemeyer}, \citenamefont {Marra},
  \citenamefont {Jiang}, \citenamefont {Lichtman}, \citenamefont {Sun},
  \citenamefont {Ebert},\ and\ \citenamefont {Saffman}}]{Saffman2019arXiv}%
  \BibitemOpen
  \bibfield  {author} {\bibinfo {author} {\bibfnamefont {T.~M.}\ \bibnamefont
  {Graham}}, \bibinfo {author} {\bibfnamefont {M.}~\bibnamefont {Kwon}},
  \bibinfo {author} {\bibfnamefont {B.}~\bibnamefont {Grinkemeyer}}, \bibinfo
  {author} {\bibfnamefont {Z.}~\bibnamefont {Marra}}, \bibinfo {author}
  {\bibfnamefont {X.}~\bibnamefont {Jiang}}, \bibinfo {author} {\bibfnamefont
  {M.~T.}\ \bibnamefont {Lichtman}}, \bibinfo {author} {\bibfnamefont
  {Y.}~\bibnamefont {Sun}}, \bibinfo {author} {\bibfnamefont {M.}~\bibnamefont
  {Ebert}},\ and\ \bibinfo {author} {\bibfnamefont {M.}~\bibnamefont
  {Saffman}},\ }\href {https://arxiv.org/abs/1908.06103} {\bibinfo  {journal}
  {arXiv:1908.06103}\ }\BibitemShut {NoStop}%
\bibitem [{\citenamefont {Omran}\ \emph {et~al.}(2019)\citenamefont {Omran},
  \citenamefont {Levine}, \citenamefont {Keesling}, \citenamefont {Semeghini},
  \citenamefont {Wang}, \citenamefont {Ebadi}, \citenamefont {Bernien},
  \citenamefont {Zibrov}, \citenamefont {Pichler}, \citenamefont {Choi},
  \citenamefont {Cui}, \citenamefont {Rossignolo}, \citenamefont {Rembold},
  \citenamefont {Montangero}, \citenamefont {Calarco}, \citenamefont {Endres},
  \citenamefont {Greiner}, \citenamefont {Vuleti{\'c}},\ and\ \citenamefont
  {Lukin}}]{Omran570}%
  \BibitemOpen
\bibfield  {journal} {  }\bibfield  {author} {\bibinfo {author} {\bibfnamefont
  {A.}~\bibnamefont {Omran}}, \bibinfo {author} {\bibfnamefont
  {H.}~\bibnamefont {Levine}}, \bibinfo {author} {\bibfnamefont
  {A.}~\bibnamefont {Keesling}}, \bibinfo {author} {\bibfnamefont
  {G.}~\bibnamefont {Semeghini}}, \bibinfo {author} {\bibfnamefont {T.~T.}\
  \bibnamefont {Wang}}, \bibinfo {author} {\bibfnamefont {S.}~\bibnamefont
  {Ebadi}}, \bibinfo {author} {\bibfnamefont {H.}~\bibnamefont {Bernien}},
  \bibinfo {author} {\bibfnamefont {A.~S.}\ \bibnamefont {Zibrov}}, \bibinfo
  {author} {\bibfnamefont {H.}~\bibnamefont {Pichler}}, \bibinfo {author}
  {\bibfnamefont {S.}~\bibnamefont {Choi}}, \bibinfo {author} {\bibfnamefont
  {J.}~\bibnamefont {Cui}}, \bibinfo {author} {\bibfnamefont {M.}~\bibnamefont
  {Rossignolo}}, \bibinfo {author} {\bibfnamefont {P.}~\bibnamefont {Rembold}},
  \bibinfo {author} {\bibfnamefont {S.}~\bibnamefont {Montangero}}, \bibinfo
  {author} {\bibfnamefont {T.}~\bibnamefont {Calarco}}, \bibinfo {author}
  {\bibfnamefont {M.}~\bibnamefont {Endres}}, \bibinfo {author} {\bibfnamefont
  {M.}~\bibnamefont {Greiner}}, \bibinfo {author} {\bibfnamefont
  {V.}~\bibnamefont {Vuleti{\'c}}},\ and\ \bibinfo {author} {\bibfnamefont
  {M.~D.}\ \bibnamefont {Lukin}},\ }\href
  {https://doi.org/10.1126/science.aax9743} {\bibfield  {journal} {\bibinfo
  {journal} {Science}\ }\textbf {\bibinfo {volume} {365}},\ \bibinfo {pages}
  {570} (\bibinfo {year} {2019})}\BibitemShut {NoStop}%
\bibitem [{\citenamefont {Levine}\ \emph {et~al.}()\citenamefont {Levine},
  \citenamefont {Keesling}, \citenamefont {Semeghini}, \citenamefont {Omran},
  \citenamefont {Wang}, \citenamefont {Ebadi}, \citenamefont {Bernien},
  \citenamefont {Greiner}, \citenamefont {Vuleti{\'c}}, \citenamefont
  {Pichler},\ and\ \citenamefont {Lukin}}]{Lukin2019arXiv}%
  \BibitemOpen
  \bibfield  {author} {\bibinfo {author} {\bibfnamefont {H.}~\bibnamefont
  {Levine}}, \bibinfo {author} {\bibfnamefont {A.}~\bibnamefont {Keesling}},
  \bibinfo {author} {\bibfnamefont {G.}~\bibnamefont {Semeghini}}, \bibinfo
  {author} {\bibfnamefont {A.}~\bibnamefont {Omran}}, \bibinfo {author}
  {\bibfnamefont {T.~T.}\ \bibnamefont {Wang}}, \bibinfo {author}
  {\bibfnamefont {S.}~\bibnamefont {Ebadi}}, \bibinfo {author} {\bibfnamefont
  {H.}~\bibnamefont {Bernien}}, \bibinfo {author} {\bibfnamefont
  {M.}~\bibnamefont {Greiner}}, \bibinfo {author} {\bibfnamefont
  {V.}~\bibnamefont {Vuleti{\'c}}}, \bibinfo {author} {\bibfnamefont
  {H.}~\bibnamefont {Pichler}},\ and\ \bibinfo {author} {\bibfnamefont {M.~D.}\
  \bibnamefont {Lukin}},\ }\href {https://arxiv.org/abs/1908.06101} {\bibinfo
  {journal} {arXiv:1908.06101}\ }\BibitemShut {NoStop}%
\bibitem [{\citenamefont {Saffman}\ and\ \citenamefont
  {Walker}(2002)}]{PhysRevA.66.065403}%
  \BibitemOpen
\bibfield  {journal} {  }\bibfield  {author} {\bibinfo {author} {\bibfnamefont
  {M.}~\bibnamefont {Saffman}}\ and\ \bibinfo {author} {\bibfnamefont {T.~G.}\
  \bibnamefont {Walker}},\ }\href {https://doi.org/10.1103/PhysRevA.66.065403}
  {\bibfield  {journal} {\bibinfo  {journal} {Phys. Rev. A}\ }\textbf {\bibinfo
  {volume} {66}},\ \bibinfo {pages} {065403} (\bibinfo {year}
  {2002})}\BibitemShut {NoStop}%
\bibitem [{\citenamefont {Brion}\ \emph {et~al.}(2007)\citenamefont {Brion},
  \citenamefont {M\o{}lmer},\ and\ \citenamefont
  {Saffman}}]{PhysRevLett.99.260501}%
  \BibitemOpen
  \bibfield  {author} {\bibinfo {author} {\bibfnamefont {E.}~\bibnamefont
  {Brion}}, \bibinfo {author} {\bibfnamefont {K.}~\bibnamefont {M\o{}lmer}},\
  and\ \bibinfo {author} {\bibfnamefont {M.}~\bibnamefont {Saffman}},\ }\href
  {https://doi.org/10.1103/PhysRevLett.99.260501} {\bibfield  {journal}
  {\bibinfo  {journal} {Phys. Rev. Lett.}\ }\textbf {\bibinfo {volume} {99}},\
  \bibinfo {pages} {260501} (\bibinfo {year} {2007})}\BibitemShut {NoStop}%
\bibitem [{\citenamefont {Honer}\ \emph {et~al.}(2011)\citenamefont {Honer},
  \citenamefont {L\"ow}, \citenamefont {Weimer}, \citenamefont {Pfau},\ and\
  \citenamefont {B\"uchler}}]{PhysRevLett.107.093601}%
  \BibitemOpen
  \bibfield  {author} {\bibinfo {author} {\bibfnamefont {J.}~\bibnamefont
  {Honer}}, \bibinfo {author} {\bibfnamefont {R.}~\bibnamefont {L\"ow}},
  \bibinfo {author} {\bibfnamefont {H.}~\bibnamefont {Weimer}}, \bibinfo
  {author} {\bibfnamefont {T.}~\bibnamefont {Pfau}},\ and\ \bibinfo {author}
  {\bibfnamefont {H.~P.}\ \bibnamefont {B\"uchler}},\ }\href
  {https://doi.org/10.1103/PhysRevLett.107.093601} {\bibfield  {journal}
  {\bibinfo  {journal} {Phys. Rev. Lett.}\ }\textbf {\bibinfo {volume} {107}},\
  \bibinfo {pages} {093601} (\bibinfo {year} {2011})}\BibitemShut {NoStop}%
\bibitem [{\citenamefont {Gorshkov}\ \emph {et~al.}(2011)\citenamefont
  {Gorshkov}, \citenamefont {Otterbach}, \citenamefont {Fleischhauer},
  \citenamefont {Pohl},\ and\ \citenamefont {Lukin}}]{PhysRevLett.107.133602}%
  \BibitemOpen
  \bibfield  {author} {\bibinfo {author} {\bibfnamefont {A.~V.}\ \bibnamefont
  {Gorshkov}}, \bibinfo {author} {\bibfnamefont {J.}~\bibnamefont {Otterbach}},
  \bibinfo {author} {\bibfnamefont {M.}~\bibnamefont {Fleischhauer}}, \bibinfo
  {author} {\bibfnamefont {T.}~\bibnamefont {Pohl}},\ and\ \bibinfo {author}
  {\bibfnamefont {M.~D.}\ \bibnamefont {Lukin}},\ }\href
  {https://doi.org/10.1103/PhysRevLett.107.133602} {\bibfield  {journal}
  {\bibinfo  {journal} {Phys. Rev. Lett.}\ }\textbf {\bibinfo {volume} {107}},\
  \bibinfo {pages} {133602} (\bibinfo {year} {2011})}\BibitemShut {NoStop}%
\bibitem [{\citenamefont {Parigi}\ \emph {et~al.}(2012)\citenamefont {Parigi},
  \citenamefont {Bimbard}, \citenamefont {Stanojevic}, \citenamefont
  {Hilliard}, \citenamefont {Nogrette}, \citenamefont {Tualle-Brouri},
  \citenamefont {Ourjoumtsev},\ and\ \citenamefont
  {Grangier}}]{PhysRevLett.109.233602}%
  \BibitemOpen
  \bibfield  {author} {\bibinfo {author} {\bibfnamefont {V.}~\bibnamefont
  {Parigi}}, \bibinfo {author} {\bibfnamefont {E.}~\bibnamefont {Bimbard}},
  \bibinfo {author} {\bibfnamefont {J.}~\bibnamefont {Stanojevic}}, \bibinfo
  {author} {\bibfnamefont {A.~J.}\ \bibnamefont {Hilliard}}, \bibinfo {author}
  {\bibfnamefont {F.}~\bibnamefont {Nogrette}}, \bibinfo {author}
  {\bibfnamefont {R.}~\bibnamefont {Tualle-Brouri}}, \bibinfo {author}
  {\bibfnamefont {A.}~\bibnamefont {Ourjoumtsev}},\ and\ \bibinfo {author}
  {\bibfnamefont {P.}~\bibnamefont {Grangier}},\ }\href
  {https://doi.org/10.1103/PhysRevLett.109.233602} {\bibfield  {journal}
  {\bibinfo  {journal} {Phys. Rev. Lett.}\ }\textbf {\bibinfo {volume} {109}},\
  \bibinfo {pages} {233602} (\bibinfo {year} {2012})}\BibitemShut {NoStop}%
\bibitem [{\citenamefont {Dudin}\ and\ \citenamefont
  {Kuzmich}(2012)}]{Dudin887Science}%
  \BibitemOpen
  \bibfield  {author} {\bibinfo {author} {\bibfnamefont {Y.~O.}\ \bibnamefont
  {Dudin}}\ and\ \bibinfo {author} {\bibfnamefont {A.}~\bibnamefont
  {Kuzmich}},\ }\href {https://doi.org/10.1126/science.1217901} {\bibfield
  {journal} {\bibinfo  {journal} {Science}\ }\textbf {\bibinfo {volume}
  {336}},\ \bibinfo {pages} {887} (\bibinfo {year} {2012})}\BibitemShut
  {NoStop}%
\bibitem [{\citenamefont {Maxwell}\ \emph {et~al.}(2013)\citenamefont
  {Maxwell}, \citenamefont {Szwer}, \citenamefont {Paredes-Barato},
  \citenamefont {Busche}, \citenamefont {Pritchard}, \citenamefont {Gauguet},
  \citenamefont {Weatherill}, \citenamefont {Jones},\ and\ \citenamefont
  {Adams}}]{PhysRevLett.110.103001}%
  \BibitemOpen
  \bibfield  {author} {\bibinfo {author} {\bibfnamefont {D.}~\bibnamefont
  {Maxwell}}, \bibinfo {author} {\bibfnamefont {D.~J.}\ \bibnamefont {Szwer}},
  \bibinfo {author} {\bibfnamefont {D.}~\bibnamefont {Paredes-Barato}},
  \bibinfo {author} {\bibfnamefont {H.}~\bibnamefont {Busche}}, \bibinfo
  {author} {\bibfnamefont {J.~D.}\ \bibnamefont {Pritchard}}, \bibinfo {author}
  {\bibfnamefont {A.}~\bibnamefont {Gauguet}}, \bibinfo {author} {\bibfnamefont
  {K.~J.}\ \bibnamefont {Weatherill}}, \bibinfo {author} {\bibfnamefont
  {M.~P.~A.}\ \bibnamefont {Jones}},\ and\ \bibinfo {author} {\bibfnamefont
  {C.~S.}\ \bibnamefont {Adams}},\ }\href
  {https://doi.org/10.1103/PhysRevLett.110.103001} {\bibfield  {journal}
  {\bibinfo  {journal} {Phys. Rev. Lett.}\ }\textbf {\bibinfo {volume} {110}},\
  \bibinfo {pages} {103001} (\bibinfo {year} {2013})}\BibitemShut {NoStop}%
\bibitem [{\citenamefont {Gorniaczyk}\ \emph {et~al.}(2014)\citenamefont
  {Gorniaczyk}, \citenamefont {Tresp}, \citenamefont {Schmidt}, \citenamefont
  {Fedder},\ and\ \citenamefont {Hofferberth}}]{PhysRevLett.113.053601}%
  \BibitemOpen
  \bibfield  {author} {\bibinfo {author} {\bibfnamefont {H.}~\bibnamefont
  {Gorniaczyk}}, \bibinfo {author} {\bibfnamefont {C.}~\bibnamefont {Tresp}},
  \bibinfo {author} {\bibfnamefont {J.}~\bibnamefont {Schmidt}}, \bibinfo
  {author} {\bibfnamefont {H.}~\bibnamefont {Fedder}},\ and\ \bibinfo {author}
  {\bibfnamefont {S.}~\bibnamefont {Hofferberth}},\ }\href
  {https://doi.org/10.1103/PhysRevLett.113.053601} {\bibfield  {journal}
  {\bibinfo  {journal} {Phys. Rev. Lett.}\ }\textbf {\bibinfo {volume} {113}},\
  \bibinfo {pages} {053601} (\bibinfo {year} {2014})}\BibitemShut {NoStop}%
\bibitem [{\citenamefont {Tresp}\ \emph {et~al.}(2016)\citenamefont {Tresp},
  \citenamefont {Zimmer}, \citenamefont {Mirgorodskiy}, \citenamefont
  {Gorniaczyk}, \citenamefont {Paris-Mandoki},\ and\ \citenamefont
  {Hofferberth}}]{PhysRevLett.117.223001}%
  \BibitemOpen
  \bibfield  {author} {\bibinfo {author} {\bibfnamefont {C.}~\bibnamefont
  {Tresp}}, \bibinfo {author} {\bibfnamefont {C.}~\bibnamefont {Zimmer}},
  \bibinfo {author} {\bibfnamefont {I.}~\bibnamefont {Mirgorodskiy}}, \bibinfo
  {author} {\bibfnamefont {H.}~\bibnamefont {Gorniaczyk}}, \bibinfo {author}
  {\bibfnamefont {A.}~\bibnamefont {Paris-Mandoki}},\ and\ \bibinfo {author}
  {\bibfnamefont {S.}~\bibnamefont {Hofferberth}},\ }\href
  {https://doi.org/10.1103/PhysRevLett.117.223001} {\bibfield  {journal}
  {\bibinfo  {journal} {Phys. Rev. Lett.}\ }\textbf {\bibinfo {volume} {117}},\
  \bibinfo {pages} {223001} (\bibinfo {year} {2016})}\BibitemShut {NoStop}%
\bibitem [{\citenamefont {Hao}\ \emph {et~al.}(2015)\citenamefont {Hao},
  \citenamefont {Lin}, \citenamefont {Xia}, \citenamefont {Lin}, \citenamefont
  {Niu},\ and\ \citenamefont {Gong}}]{Hao2015srep}%
  \BibitemOpen
  \bibfield  {author} {\bibinfo {author} {\bibfnamefont {Y.~M.}\ \bibnamefont
  {Hao}}, \bibinfo {author} {\bibfnamefont {G.~W.}\ \bibnamefont {Lin}},
  \bibinfo {author} {\bibfnamefont {K.}~\bibnamefont {Xia}}, \bibinfo {author}
  {\bibfnamefont {X.~M.}\ \bibnamefont {Lin}}, \bibinfo {author} {\bibfnamefont
  {Y.~P.}\ \bibnamefont {Niu}},\ and\ \bibinfo {author} {\bibfnamefont {S.~Q.}\
  \bibnamefont {Gong}},\ }\href {https://doi.org/10.1038/srep10005} {\bibfield
  {journal} {\bibinfo  {journal} {Sci. Rep.}\ }\textbf {\bibinfo {volume}
  {5}},\ \bibinfo {pages} {10005} (\bibinfo {year} {2015})}\BibitemShut
  {NoStop}%
\bibitem [{\citenamefont {Das}\ \emph {et~al.}(2016)\citenamefont {Das},
  \citenamefont {Grankin}, \citenamefont {Iakoupov}, \citenamefont {Brion},
  \citenamefont {Borregaard}, \citenamefont {Boddeda}, \citenamefont {Usmani},
  \citenamefont {Ourjoumtsev}, \citenamefont {Grangier},\ and\ \citenamefont
  {S\o{}rensen}}]{PhysRevA.93.040303}%
  \BibitemOpen
  \bibfield  {author} {\bibinfo {author} {\bibfnamefont {S.}~\bibnamefont
  {Das}}, \bibinfo {author} {\bibfnamefont {A.}~\bibnamefont {Grankin}},
  \bibinfo {author} {\bibfnamefont {I.}~\bibnamefont {Iakoupov}}, \bibinfo
  {author} {\bibfnamefont {E.}~\bibnamefont {Brion}}, \bibinfo {author}
  {\bibfnamefont {J.}~\bibnamefont {Borregaard}}, \bibinfo {author}
  {\bibfnamefont {R.}~\bibnamefont {Boddeda}}, \bibinfo {author} {\bibfnamefont
  {I.}~\bibnamefont {Usmani}}, \bibinfo {author} {\bibfnamefont
  {A.}~\bibnamefont {Ourjoumtsev}}, \bibinfo {author} {\bibfnamefont
  {P.}~\bibnamefont {Grangier}},\ and\ \bibinfo {author} {\bibfnamefont
  {A.~S.}\ \bibnamefont {S\o{}rensen}},\ }\href
  {https://doi.org/10.1103/PhysRevA.93.040303} {\bibfield  {journal} {\bibinfo
  {journal} {Phys. Rev. A}\ }\textbf {\bibinfo {volume} {93}},\ \bibinfo
  {pages} {040303} (\bibinfo {year} {2016})}\BibitemShut {NoStop}%
\bibitem [{\citenamefont {Ningyuan}\ \emph {et~al.}(2016)\citenamefont
  {Ningyuan}, \citenamefont {Georgakopoulos}, \citenamefont {Ryou},
  \citenamefont {Schine}, \citenamefont {Sommer},\ and\ \citenamefont
  {Simon}}]{PhysRevA.93.041802}%
  \BibitemOpen
  \bibfield  {author} {\bibinfo {author} {\bibfnamefont {J.}~\bibnamefont
  {Ningyuan}}, \bibinfo {author} {\bibfnamefont {A.}~\bibnamefont
  {Georgakopoulos}}, \bibinfo {author} {\bibfnamefont {A.}~\bibnamefont
  {Ryou}}, \bibinfo {author} {\bibfnamefont {N.}~\bibnamefont {Schine}},
  \bibinfo {author} {\bibfnamefont {A.}~\bibnamefont {Sommer}},\ and\ \bibinfo
  {author} {\bibfnamefont {J.}~\bibnamefont {Simon}},\ }\href
  {https://doi.org/10.1103/PhysRevA.93.041802} {\bibfield  {journal} {\bibinfo
  {journal} {Phys. Rev. A}\ }\textbf {\bibinfo {volume} {93}},\ \bibinfo
  {pages} {041802} (\bibinfo {year} {2016})}\BibitemShut {NoStop}%
\bibitem [{\citenamefont {Wade}\ \emph {et~al.}(2016)\citenamefont {Wade},
  \citenamefont {Mattioli},\ and\ \citenamefont
  {M\o{}lmer}}]{PhysRevA.94.053830}%
  \BibitemOpen
  \bibfield  {author} {\bibinfo {author} {\bibfnamefont {A.~C.~J.}\
  \bibnamefont {Wade}}, \bibinfo {author} {\bibfnamefont {M.}~\bibnamefont
  {Mattioli}},\ and\ \bibinfo {author} {\bibfnamefont {K.}~\bibnamefont
  {M\o{}lmer}},\ }\href {https://doi.org/10.1103/PhysRevA.94.053830} {\bibfield
   {journal} {\bibinfo  {journal} {Phys. Rev. A}\ }\textbf {\bibinfo {volume}
  {94}},\ \bibinfo {pages} {053830} (\bibinfo {year} {2016})}\BibitemShut
  {NoStop}%
\bibitem [{\citenamefont {Lee}\ \emph {et~al.}(2017)\citenamefont {Lee},
  \citenamefont {Martin}, \citenamefont {Jau}, \citenamefont {Keating},
  \citenamefont {Deutsch},\ and\ \citenamefont
  {Biedermann}}]{PhysRevA.95.041801}%
  \BibitemOpen
  \bibfield  {author} {\bibinfo {author} {\bibfnamefont {J.}~\bibnamefont
  {Lee}}, \bibinfo {author} {\bibfnamefont {M.~J.}\ \bibnamefont {Martin}},
  \bibinfo {author} {\bibfnamefont {Y.-Y.}\ \bibnamefont {Jau}}, \bibinfo
  {author} {\bibfnamefont {T.}~\bibnamefont {Keating}}, \bibinfo {author}
  {\bibfnamefont {I.~H.}\ \bibnamefont {Deutsch}},\ and\ \bibinfo {author}
  {\bibfnamefont {G.~W.}\ \bibnamefont {Biedermann}},\ }\href
  {https://doi.org/10.1103/PhysRevA.95.041801} {\bibfield  {journal} {\bibinfo
  {journal} {Phys. Rev. A}\ }\textbf {\bibinfo {volume} {95}},\ \bibinfo
  {pages} {041801} (\bibinfo {year} {2017})}\BibitemShut {NoStop}%
\bibitem [{\citenamefont {Sun}\ and\ \citenamefont
  {Chen}(2018)}]{OPTICA.5.001492}%
  \BibitemOpen
  \bibfield  {author} {\bibinfo {author} {\bibfnamefont {Y.}~\bibnamefont
  {Sun}}\ and\ \bibinfo {author} {\bibfnamefont {P.-X.}\ \bibnamefont {Chen}},\
  }\href {https://doi.org/10.1364/OPTICA.5.001492} {\bibfield  {journal}
  {\bibinfo  {journal} {Optica}\ }\textbf {\bibinfo {volume} {5}},\ \bibinfo
  {pages} {1492} (\bibinfo {year} {2018})}\BibitemShut {NoStop}%
\bibitem [{\citenamefont {Paredes-Barato}\ and\ \citenamefont
  {Adams}(2014)}]{PhysRevLett.112.040501}%
  \BibitemOpen
  \bibfield  {author} {\bibinfo {author} {\bibfnamefont {D.}~\bibnamefont
  {Paredes-Barato}}\ and\ \bibinfo {author} {\bibfnamefont {C.~S.}\
  \bibnamefont {Adams}},\ }\href
  {https://doi.org/10.1103/PhysRevLett.112.040501} {\bibfield  {journal}
  {\bibinfo  {journal} {Phys. Rev. Lett.}\ }\textbf {\bibinfo {volume} {112}},\
  \bibinfo {pages} {040501} (\bibinfo {year} {2014})}\BibitemShut {NoStop}%
\bibitem [{\citenamefont {Ebert}\ \emph {et~al.}(2015)\citenamefont {Ebert},
  \citenamefont {Kwon}, \citenamefont {Walker},\ and\ \citenamefont
  {Saffman}}]{PhysRevLett.115.093601}%
  \BibitemOpen
  \bibfield  {author} {\bibinfo {author} {\bibfnamefont {M.}~\bibnamefont
  {Ebert}}, \bibinfo {author} {\bibfnamefont {M.}~\bibnamefont {Kwon}},
  \bibinfo {author} {\bibfnamefont {T.~G.}\ \bibnamefont {Walker}},\ and\
  \bibinfo {author} {\bibfnamefont {M.}~\bibnamefont {Saffman}},\ }\href
  {https://doi.org/10.1103/PhysRevLett.115.093601} {\bibfield  {journal}
  {\bibinfo  {journal} {Phys. Rev. Lett.}\ }\textbf {\bibinfo {volume} {115}},\
  \bibinfo {pages} {093601} (\bibinfo {year} {2015})}\BibitemShut {NoStop}%
\bibitem [{\citenamefont {Lahad}\ and\ \citenamefont
  {Firstenberg}(2017)}]{PhysRevLett.119.113601}%
  \BibitemOpen
  \bibfield  {author} {\bibinfo {author} {\bibfnamefont {O.}~\bibnamefont
  {Lahad}}\ and\ \bibinfo {author} {\bibfnamefont {O.}~\bibnamefont
  {Firstenberg}},\ }\href {https://doi.org/10.1103/PhysRevLett.119.113601}
  {\bibfield  {journal} {\bibinfo  {journal} {Phys. Rev. Lett.}\ }\textbf
  {\bibinfo {volume} {119}},\ \bibinfo {pages} {113601} (\bibinfo {year}
  {2017})}\BibitemShut {NoStop}%
\bibitem [{\citenamefont {Khazali}\ \emph {et~al.}(2015)\citenamefont
  {Khazali}, \citenamefont {Heshami},\ and\ \citenamefont
  {Simon}}]{PhysRevA.91.030301}%
  \BibitemOpen
  \bibfield  {author} {\bibinfo {author} {\bibfnamefont {M.}~\bibnamefont
  {Khazali}}, \bibinfo {author} {\bibfnamefont {K.}~\bibnamefont {Heshami}},\
  and\ \bibinfo {author} {\bibfnamefont {C.}~\bibnamefont {Simon}},\ }\href
  {https://doi.org/10.1103/PhysRevA.91.030301} {\bibfield  {journal} {\bibinfo
  {journal} {Phys. Rev. A}\ }\textbf {\bibinfo {volume} {91}},\ \bibinfo
  {pages} {030301} (\bibinfo {year} {2015})}\BibitemShut {NoStop}%
\bibitem [{\citenamefont {Tiarks}\ \emph {et~al.}(2019)\citenamefont {Tiarks},
  \citenamefont {Schmidt-Eberle}, \citenamefont {Stolz}, \citenamefont
  {Rempe},\ and\ \citenamefont {Duerr}}]{ISI:000457492900011}%
  \BibitemOpen
  \bibfield  {author} {\bibinfo {author} {\bibfnamefont {D.}~\bibnamefont
  {Tiarks}}, \bibinfo {author} {\bibfnamefont {S.}~\bibnamefont
  {Schmidt-Eberle}}, \bibinfo {author} {\bibfnamefont {T.}~\bibnamefont
  {Stolz}}, \bibinfo {author} {\bibfnamefont {G.}~\bibnamefont {Rempe}},\ and\
  \bibinfo {author} {\bibfnamefont {S.}~\bibnamefont {Duerr}},\ }\href
  {https://doi.org/{10.1038/s41567-018-0313-7}} {\bibfield  {journal} {\bibinfo
   {journal} {{NATURE PHYSICS}}\ }\textbf {\bibinfo {volume} {{15}}},\ \bibinfo
  {pages} {{124}} (\bibinfo {year} {{2019}})}\BibitemShut {NoStop}%
\bibitem [{\citenamefont {Saffman}\ and\ \citenamefont
  {Walker}(2005)}]{PhysRevA.72.022347}%
  \BibitemOpen
  \bibfield  {author} {\bibinfo {author} {\bibfnamefont {M.}~\bibnamefont
  {Saffman}}\ and\ \bibinfo {author} {\bibfnamefont {T.~G.}\ \bibnamefont
  {Walker}},\ }\href {https://doi.org/10.1103/PhysRevA.72.022347} {\bibfield
  {journal} {\bibinfo  {journal} {Phys. Rev. A}\ }\textbf {\bibinfo {volume}
  {72}},\ \bibinfo {pages} {022347} (\bibinfo {year} {2005})}\BibitemShut
  {NoStop}%
\bibitem [{\citenamefont {Walker}\ and\ \citenamefont
  {Saffman}(2008)}]{PhysRevA.77.032723}%
  \BibitemOpen
  \bibfield  {author} {\bibinfo {author} {\bibfnamefont {T.~G.}\ \bibnamefont
  {Walker}}\ and\ \bibinfo {author} {\bibfnamefont {M.}~\bibnamefont
  {Saffman}},\ }\href {https://doi.org/10.1103/PhysRevA.77.032723} {\bibfield
  {journal} {\bibinfo  {journal} {Phys. Rev. A}\ }\textbf {\bibinfo {volume}
  {77}},\ \bibinfo {pages} {032723} (\bibinfo {year} {2008})}\BibitemShut
  {NoStop}%
\bibitem [{\citenamefont {Beterov}\ \emph {et~al.}(2013)\citenamefont
  {Beterov}, \citenamefont {Saffman}, \citenamefont {Yakshina}, \citenamefont
  {Zhukov}, \citenamefont {Tretyakov}, \citenamefont {Entin}, \citenamefont
  {Ryabtsev}, \citenamefont {Mansell}, \citenamefont {MacCormick},
  \citenamefont {Bergamini},\ and\ \citenamefont
  {Fedoruk}}]{PhysRevA.88.010303}%
  \BibitemOpen
  \bibfield  {author} {\bibinfo {author} {\bibfnamefont {I.~I.}\ \bibnamefont
  {Beterov}}, \bibinfo {author} {\bibfnamefont {M.}~\bibnamefont {Saffman}},
  \bibinfo {author} {\bibfnamefont {E.~A.}\ \bibnamefont {Yakshina}}, \bibinfo
  {author} {\bibfnamefont {V.~P.}\ \bibnamefont {Zhukov}}, \bibinfo {author}
  {\bibfnamefont {D.~B.}\ \bibnamefont {Tretyakov}}, \bibinfo {author}
  {\bibfnamefont {V.~M.}\ \bibnamefont {Entin}}, \bibinfo {author}
  {\bibfnamefont {I.~I.}\ \bibnamefont {Ryabtsev}}, \bibinfo {author}
  {\bibfnamefont {C.~W.}\ \bibnamefont {Mansell}}, \bibinfo {author}
  {\bibfnamefont {C.}~\bibnamefont {MacCormick}}, \bibinfo {author}
  {\bibfnamefont {S.}~\bibnamefont {Bergamini}},\ and\ \bibinfo {author}
  {\bibfnamefont {M.~P.}\ \bibnamefont {Fedoruk}},\ }\href
  {https://doi.org/10.1103/PhysRevA.88.010303} {\bibfield  {journal} {\bibinfo
  {journal} {Phys. Rev. A}\ }\textbf {\bibinfo {volume} {88}},\ \bibinfo
  {pages} {010303} (\bibinfo {year} {2013})}\BibitemShut {NoStop}%
\bibitem [{\citenamefont {Xia}\ \emph {et~al.}(2013)\citenamefont {Xia},
  \citenamefont {Zhang},\ and\ \citenamefont {Saffman}}]{PhysRevA.88.062337}%
  \BibitemOpen
  \bibfield  {author} {\bibinfo {author} {\bibfnamefont {T.}~\bibnamefont
  {Xia}}, \bibinfo {author} {\bibfnamefont {X.~L.}\ \bibnamefont {Zhang}},\
  and\ \bibinfo {author} {\bibfnamefont {M.}~\bibnamefont {Saffman}},\ }\href
  {https://doi.org/10.1103/PhysRevA.88.062337} {\bibfield  {journal} {\bibinfo
  {journal} {Phys. Rev. A}\ }\textbf {\bibinfo {volume} {88}},\ \bibinfo
  {pages} {062337} (\bibinfo {year} {2013})}\BibitemShut {NoStop}%
\bibitem [{\citenamefont {Keating}\ \emph {et~al.}(2015)\citenamefont
  {Keating}, \citenamefont {Cook}, \citenamefont {Hankin}, \citenamefont {Jau},
  \citenamefont {Biedermann},\ and\ \citenamefont
  {Deutsch}}]{PhysRevA.91.012337}%
  \BibitemOpen
  \bibfield  {author} {\bibinfo {author} {\bibfnamefont {T.}~\bibnamefont
  {Keating}}, \bibinfo {author} {\bibfnamefont {R.~L.}\ \bibnamefont {Cook}},
  \bibinfo {author} {\bibfnamefont {A.~M.}\ \bibnamefont {Hankin}}, \bibinfo
  {author} {\bibfnamefont {Y.-Y.}\ \bibnamefont {Jau}}, \bibinfo {author}
  {\bibfnamefont {G.~W.}\ \bibnamefont {Biedermann}},\ and\ \bibinfo {author}
  {\bibfnamefont {I.~H.}\ \bibnamefont {Deutsch}},\ }\href
  {https://doi.org/10.1103/PhysRevA.91.012337} {\bibfield  {journal} {\bibinfo
  {journal} {Phys. Rev. A}\ }\textbf {\bibinfo {volume} {91}},\ \bibinfo
  {pages} {012337} (\bibinfo {year} {2015})}\BibitemShut {NoStop}%
\bibitem [{\citenamefont {Beterov}\ and\ \citenamefont
  {Saffman}(2015)}]{PhysRevA.92.042710}%
  \BibitemOpen
  \bibfield  {author} {\bibinfo {author} {\bibfnamefont {I.~I.}\ \bibnamefont
  {Beterov}}\ and\ \bibinfo {author} {\bibfnamefont {M.}~\bibnamefont
  {Saffman}},\ }\href {https://doi.org/10.1103/PhysRevA.92.042710} {\bibfield
  {journal} {\bibinfo  {journal} {Phys. Rev. A}\ }\textbf {\bibinfo {volume}
  {92}},\ \bibinfo {pages} {042710} (\bibinfo {year} {2015})}\BibitemShut
  {NoStop}%
\bibitem [{\citenamefont {Theis}\ \emph {et~al.}(2016)\citenamefont {Theis},
  \citenamefont {Motzoi}, \citenamefont {Wilhelm},\ and\ \citenamefont
  {Saffman}}]{PhysRevA.94.032306}%
  \BibitemOpen
  \bibfield  {author} {\bibinfo {author} {\bibfnamefont {L.~S.}\ \bibnamefont
  {Theis}}, \bibinfo {author} {\bibfnamefont {F.}~\bibnamefont {Motzoi}},
  \bibinfo {author} {\bibfnamefont {F.~K.}\ \bibnamefont {Wilhelm}},\ and\
  \bibinfo {author} {\bibfnamefont {M.}~\bibnamefont {Saffman}},\ }\href
  {https://doi.org/10.1103/PhysRevA.94.032306} {\bibfield  {journal} {\bibinfo
  {journal} {Phys. Rev. A}\ }\textbf {\bibinfo {volume} {94}},\ \bibinfo
  {pages} {032306} (\bibinfo {year} {2016})}\BibitemShut {NoStop}%
\bibitem [{\citenamefont {Petrosyan}\ \emph {et~al.}(2017)\citenamefont
  {Petrosyan}, \citenamefont {Motzoi}, \citenamefont {Saffman},\ and\
  \citenamefont {M\o{}lmer}}]{PhysRevA.96.042306}%
  \BibitemOpen
  \bibfield  {author} {\bibinfo {author} {\bibfnamefont {D.}~\bibnamefont
  {Petrosyan}}, \bibinfo {author} {\bibfnamefont {F.}~\bibnamefont {Motzoi}},
  \bibinfo {author} {\bibfnamefont {M.}~\bibnamefont {Saffman}},\ and\ \bibinfo
  {author} {\bibfnamefont {K.}~\bibnamefont {M\o{}lmer}},\ }\href
  {https://doi.org/10.1103/PhysRevA.96.042306} {\bibfield  {journal} {\bibinfo
  {journal} {Phys. Rev. A}\ }\textbf {\bibinfo {volume} {96}},\ \bibinfo
  {pages} {042306} (\bibinfo {year} {2017})}\BibitemShut {NoStop}%
\bibitem [{\citenamefont {Sun}\ \emph {et~al.}()\citenamefont {Sun},
  \citenamefont {Xu},\ and\ \citenamefont {Chen}}]{Sun2018arXiv}%
  \BibitemOpen
  \bibfield  {author} {\bibinfo {author} {\bibfnamefont {Y.}~\bibnamefont
  {Sun}}, \bibinfo {author} {\bibfnamefont {P.}~\bibnamefont {Xu}},\ and\
  \bibinfo {author} {\bibfnamefont {P.-X.}\ \bibnamefont {Chen}},\ }\href
  {https://arxiv.org/abs/1812.03822} {\bibinfo  {journal} {arXiv:1812.03822}\
  }\BibitemShut {NoStop}%
\bibitem [{\citenamefont {Morris}\ and\ \citenamefont
  {Shore}(1983)}]{PhysRevA.27.906}%
  \BibitemOpen
\bibfield  {journal} {  }\bibfield  {author} {\bibinfo {author} {\bibfnamefont
  {J.~R.}\ \bibnamefont {Morris}}\ and\ \bibinfo {author} {\bibfnamefont
  {B.~W.}\ \bibnamefont {Shore}},\ }\href
  {https://doi.org/10.1103/PhysRevA.27.906} {\bibfield  {journal} {\bibinfo
  {journal} {Phys. Rev. A}\ }\textbf {\bibinfo {volume} {27}},\ \bibinfo
  {pages} {906} (\bibinfo {year} {1983})}\BibitemShut {NoStop}%
\bibitem [{\citenamefont {Xia}\ \emph {et~al.}(2015)\citenamefont {Xia},
  \citenamefont {Lichtman}, \citenamefont {Maller}, \citenamefont {Carr},
  \citenamefont {Piotrowicz}, \citenamefont {Isenhower},\ and\ \citenamefont
  {Saffman}}]{PhysRevLett.114.100503}%
  \BibitemOpen
  \bibfield  {author} {\bibinfo {author} {\bibfnamefont {T.}~\bibnamefont
  {Xia}}, \bibinfo {author} {\bibfnamefont {M.}~\bibnamefont {Lichtman}},
  \bibinfo {author} {\bibfnamefont {K.}~\bibnamefont {Maller}}, \bibinfo
  {author} {\bibfnamefont {A.~W.}\ \bibnamefont {Carr}}, \bibinfo {author}
  {\bibfnamefont {M.~J.}\ \bibnamefont {Piotrowicz}}, \bibinfo {author}
  {\bibfnamefont {L.}~\bibnamefont {Isenhower}},\ and\ \bibinfo {author}
  {\bibfnamefont {M.}~\bibnamefont {Saffman}},\ }\href
  {https://doi.org/10.1103/PhysRevLett.114.100503} {\bibfield  {journal}
  {\bibinfo  {journal} {Phys. Rev. Lett.}\ }\textbf {\bibinfo {volume} {114}},\
  \bibinfo {pages} {100503} (\bibinfo {year} {2015})}\BibitemShut {NoStop}%
\bibitem [{Sup()}]{SuppInfo}%
  \BibitemOpen
  \href@noop {} {\bibinfo  {journal} {See Supplemental Material at URL for more
  underlying details of waveform designing process and sub-microsecond gate
  operation}\ }\BibitemShut {NoStop}%
\bibitem [{\citenamefont {Chow}\ \emph {et~al.}(2011)\citenamefont {Chow},
  \citenamefont {C\'orcoles}, \citenamefont {Gambetta}, \citenamefont
  {Rigetti}, \citenamefont {Johnson}, \citenamefont {Smolin}, \citenamefont
  {Rozen}, \citenamefont {Keefe}, \citenamefont {Rothwell}, \citenamefont
  {Ketchen},\ and\ \citenamefont {Steffen}}]{PhysRevLett.107.080502}%
  \BibitemOpen
\bibfield  {journal} {  }\bibfield  {author} {\bibinfo {author} {\bibfnamefont
  {J.~M.}\ \bibnamefont {Chow}}, \bibinfo {author} {\bibfnamefont {A.~D.}\
  \bibnamefont {C\'orcoles}}, \bibinfo {author} {\bibfnamefont {J.~M.}\
  \bibnamefont {Gambetta}}, \bibinfo {author} {\bibfnamefont {C.}~\bibnamefont
  {Rigetti}}, \bibinfo {author} {\bibfnamefont {B.~R.}\ \bibnamefont
  {Johnson}}, \bibinfo {author} {\bibfnamefont {J.~A.}\ \bibnamefont {Smolin}},
  \bibinfo {author} {\bibfnamefont {J.~R.}\ \bibnamefont {Rozen}}, \bibinfo
  {author} {\bibfnamefont {G.~A.}\ \bibnamefont {Keefe}}, \bibinfo {author}
  {\bibfnamefont {M.~B.}\ \bibnamefont {Rothwell}}, \bibinfo {author}
  {\bibfnamefont {M.~B.}\ \bibnamefont {Ketchen}},\ and\ \bibinfo {author}
  {\bibfnamefont {M.}~\bibnamefont {Steffen}},\ }\href
  {https://doi.org/10.1103/PhysRevLett.107.080502} {\bibfield  {journal}
  {\bibinfo  {journal} {Phys. Rev. Lett.}\ }\textbf {\bibinfo {volume} {107}},\
  \bibinfo {pages} {080502} (\bibinfo {year} {2011})}\BibitemShut {NoStop}%
\bibitem [{\citenamefont {Kwon}\ \emph {et~al.}(2017)\citenamefont {Kwon},
  \citenamefont {Ebert}, \citenamefont {Walker},\ and\ \citenamefont
  {Saffman}}]{PhysRevLett.119.180504}%
  \BibitemOpen
  \bibfield  {author} {\bibinfo {author} {\bibfnamefont {M.}~\bibnamefont
  {Kwon}}, \bibinfo {author} {\bibfnamefont {M.~F.}\ \bibnamefont {Ebert}},
  \bibinfo {author} {\bibfnamefont {T.~G.}\ \bibnamefont {Walker}},\ and\
  \bibinfo {author} {\bibfnamefont {M.}~\bibnamefont {Saffman}},\ }\href
  {https://doi.org/10.1103/PhysRevLett.119.180504} {\bibfield  {journal}
  {\bibinfo  {journal} {Phys. Rev. Lett.}\ }\textbf {\bibinfo {volume} {119}},\
  \bibinfo {pages} {180504} (\bibinfo {year} {2017})}\BibitemShut {NoStop}%
\end{thebibliography}%

\end{document}